\documentclass[prl,aps,twocolumn,showpacs,amsmath,amssymb,amsfonts,floatfix,superscriptaddress]{revtex4}
\usepackage{graphicx}
\usepackage{dcolumn}
\usepackage{bm}
\usepackage[hypertex]{hyperref}
\usepackage{latexsym}
\usepackage{float}
\usepackage{supertabular}
\usepackage{longtable}
\usepackage{soul}

\begin{document}
\title{Fixation in Evolutionary Games under Non-Vanishing Selection}
\author{Mauro Mobilia} \affiliation{Department of Applied Mathematics, University of Leeds, Leeds LS2 9JT, United Kingdom}
\author{Michael Assaf}
\affiliation{Racah Institute of Physics, Hebrew University of
Jerusalem, Jerusalem 91904, Israel}
\begin{abstract}
One of the most striking effect of fluctuations in evolutionary game theory is  the possibility for mutants  to fixate (take over) an entire population. Here, we generalize a recent WKB-based theory to study fixation in evolutionary games under non-vanishing selection, and investigate the relation between selection intensity $w$ and demographic (random) fluctuations. This allows the accurate treatment of large fluctuations and yields the probability and mean times of fixation  \textit{beyond} the weak selection limit.
The power of the theory is demonstrated on prototypical models of cooperation dilemmas with multiple absorbing states.
Our predictions compare excellently with numerical simulations and, for {\it finite} $w$, significantly improve over those of the Fokker-Planck approximation.
\end{abstract}

\maketitle

\section{Introduction}
Evolutionary game theory (EGT) provides a natural theoretical framework to describe the dynamics
of systems  where successful types or behaviors, as those arising in biology, ecology and economics~\cite{Hofbauer,Nowak}, are copied by imitation and spread.
Evolutionary stability is a crucial concept in  EGT and specifies under which circumstances
a population is proof against  invasion from mutants~\cite{Hofbauer,Nowak}.
This notion was shown to be altered by finite-size fluctuations  and led to the key concept  of {\it evolutionary stability in finite populations}~\cite{Nowak}.
The latter is closely related to the notion of {\it fixation}~\cite{Nowak,Hauert-preprint}, referring to the possibility for  mutants  to take over (fixate) an entire population of wild species individuals.
Furthermore, evolutionary dynamics is characterized by the interplay between random fluctuations~\cite{Kimura} and selection,
that underlies adaptation in terms of the different reproduction potential (fitness) of the individuals. Thus, a parameter was introduced to measure the selection intensity~\cite{Nowak}.
In this context, the fixation probability of a species
 has been calculated for a finite two-species population in the weak selection limit of vanishingly small selection intensity~\cite{Nowak,Traulsen,Hauert-preprint}.
This limit is often biologically relevant and greatly simplifies the analysis
(treating selection
as a linear perturbation). However,
the behaviors obtained under strong and weak selection are often
qualitatively different (see e.g.~\cite{Antal,Traulsen}).

In this Letter, we study fixation under non-vanishing selection in EGT and provide a comprehensive analysis of the combined influence  of non-vanishing selection and random fluctuations. As exact results for  the fixation probability and mean fixation  times (MFTs) are rarely available and often unwieldy (see e.g.~\cite{Gardiner,Nowak,Hauert-preprint}), our analysis relies on the WKB (Wentzel-Kramers-Brillouin) approximation method~\cite{landaubender} directly applied to the underlying master equation~\cite{kubodykman}. This technique was recently used to treat generic birth-death systems that undergo metastable switching or extinction~\cite{AM,EK}. Importantly,
here we generalize the WKB formalism to systems with \textit{multiple} absorbing states. This theory accurately accounts for the large fluctuations not aptly captured~\cite{gaveau,Assaf} by the Fokker-Planck approximation (FPA)~\cite{Gardiner}.
We illustrate our method on two classes of prototypical models of cooperation dilemmas, the anti- coordination and coordination games,
where a coexistence state separates two absorbing states in which the population is composed of only the fixated species while the other goes extinct~\cite{Hofbauer,Nowak}. We compute  the fixation probabilities, the MFTs, as well as the complete probability distribution function
(PDF) of population sizes, and show that our theory is superior to the FPA for finite selection strength.

\section{The models}
In EGT, the fitness, or reproduction potential of an individual, is determined by the
outcome, called payoff,  of its interaction with the others as prescribed by the underlying game~\cite{Hofbauer}. In fact, when two $\textsf{A}-$individuals interact, both receive a payoff $a$. If an individual of type $\textsf{A}$
interacts with another of type $\textsf{B}$, the former receives $b$ while the latter gets a payoff $c$. Similarly, when two $\textsf{B}-$individuals interact, both get a payoff $d$. Now, assume that in a population of size $N$ there are $n$
individuals of type
$\textsf{A}$ (``mutants'') and
$N-n$ of type $\textsf{B}$ (``wild type''). The respective average payoffs (per individual) are $\Pi_{A}(n)=(n/N) a + \left[(N-n)/N\right]b\,$ and $\Pi_{B}(n)=(n/N) c + \left[(N-n)/N\right]d$~\cite{rates}, while the population mean payoff is $\bar{\Pi}(n)=\left[n \Pi_A(n)  + (N-n)\Pi_B(n) \right]/N$. For infinite ($N\to \infty$) and well-mixed
populations, the density $x\equiv n/N$ of the $\textsf{A}$ species changes according to its relative payoff and obeys the
replicator dynamics, given by the rate equation~\cite{Hofbauer,Nowak}
\begin{equation}\label{RD}
\dot{x}=x(\Pi_{A}-\bar{\Pi}).
\end{equation}
Here, we are particularly interested in anti-coordination games (ACG), where $c>a$ and $b>d$, and  in coordination games (CG), where $a>c$ and $d>b$.
In addition to the absorbing states $n=0$  and $n=N$, ACG and CG admit an interior fixed point associated with the \textit{coexistence} of $\textsf{A}$ and $\textsf{B}$ species at a density $x^*=(d-b)/(a-b-c+d)$ of $\textsf{A}$'s. According to the rate equation~(\ref{RD}), $x^*$ is an attractor in
 ACG and a repellor in CG, whereas  $x=0$ and $x=1$ are repelling  fixed points in ACG and attracting  in CG.

To account for fluctuations arising when the population size is finite,
the evolutionary dynamics is implemented in terms of fitness-dependent birth-death processes~\cite{Nowak,Hauert-preprint} describing, e.g., the evolution of the probability $P_n(t)$ to have $n$ individuals of type $\textsf{A}$ at time $t$:
\begin{equation}\label{master}
\frac{dP_n(t)}{dt}= T^+_{n-1}P_{n-1}+T^-_{n+1} P_{n+1}-[T^+_{n}+T^-_{n}]P_n.
\end{equation}
Here, an individual chosen proportionally to its fitness produces an identical offspring which replaces a randomly chosen individual~\cite{PopGen}, and the total population size $N$ is conserved. Thus, in the master equation (\ref{master}), the reaction rates for the birth/death transitions
$n \to n\pm 1$
are given by $T^{\pm}_n=\chi^{\pm}(f_A(n),f_B(n))\,n(N-n)/N^2$, where  $\chi^{\pm}(n)$
are functions of the fitness of each species, $f_A(n)=1-w +w\Pi_A(n)$ and $f_B(n)=1-w +w\Pi_B(n)$.
As often in EGT, we focus on systems evolving according to the fitness-dependent \textit{Moran process} (fMP) for which $\chi^+(n)=f_A[(n/N)f_A
+(1-n/N)f_B]^{-1}$ and $\chi^-(n)=f_B[(n/N)f_A +(1-n/N)f_B]^{-1}$~\cite{PopGen,Nowak}.
It is worth noticing that $\chi^+(n)$ and $\chi^-(n)$ intersect  only at the fixed point value $n=Nx^*$ for
 $0\leq n\leq N$, which ensures that the properties of the replicator dynamics~(\ref{RD}) are recovered
when $N\to \infty$~\cite{Hofbauer,Nowak,Hauert-preprint,Claussen06,choice}.

The fitnesses $f_A(n)$ and $f_B(n)$ are comprised of a baseline contribution  [the constant $(1-w)$] and a term accounting for selection [$w\Pi_A$ for $f_A$], where the parameter $0\leq w\leq 1$ measures the selection intensity~\cite{Nowak,Hauert-preprint}.
The latter is weak for $w\to 0$, when $T^{\pm}_n \propto \,n(N-n)/N^2$,  and  strong for $w \to 1$, when the baseline fitness becomes negligible.
As  $n\in[0,N]$ and $n=0,N$ are absorbing, the boundary conditions to Eq.~(\ref{master}) are  $T^{\pm}_0=T^{\pm}_N=0$.

\section{WKB theory of anti-coordination games}
Our WKB-based approach is presented in the framework of
ACG (\textit{e.g.} snowdrift and hawk-dove games~\cite{Hofbauer}), where the absorbing states $n=0$ or $x=0$ (all $\textsf{B}'s$), and $n=N$ or $x=1$ (all $\textsf{A}'s$)
are separated by the interior attractor $x^*$ [in the language of the rate equation~(\ref{RD})]. However, in the presence of noise $x^*$ becomes {\it metastable}, which is very naturally accounted by our theory. For $Nx^*\gg 1$, after a short relaxation time $t_r$, the system settles into a long-lived metastable state
whose population size distribution is peaked in the vicinity of $Nx^*$~\cite{Assaf}.
This implies that fixation of either  species occurs only in the aftermath of a long-lasting coexistence.
At $t\gg t_r$, only the first excited eigenvector of  (\ref{master}), $\pi_n$, called the quasi-stationary distribution (QSD), has not decayed and hence determines the shape of the metastable PDF. Indeed, at $t\gg t_r$ the higher eigenmodes in the spectral expansion of $P_n(t)$ have already decayed, and the metastable dynamics of the  population sizes PDF satisfies~\cite{Assaf}
\begin{equation} \label{pin}
P_{n}(t)\simeq \pi_n e^{-t/\tau}\quad \text{for} \quad n\in[1,N-1], 
\end{equation}
where $\sum_n\pi_n=1$. Thus, at $t\gg t_r$ the dynamics of the probabilities to be absorbed at $n=0$ and $n=N$ satisfies
\begin{eqnarray}\label{p0n}
P_{0}(t)\simeq \phi (1- e^{-t/\tau})\;,\;\;P_{N}(t) \simeq (1-\phi) (1- e^{-t/\tau}).
\end{eqnarray}
Here, $\phi^B=\phi$ and $\phi^A=1-\phi$ are the fixation probabilities of the $\textsf{B}$ and $\textsf{A}$ species, respectively, $\tau$ is the unconditional MFT, and a very strong inequality $\tau\ggg t_r$ holds. The fixation probability and MFT
are determined by the  fluxes into the absorbing states. Therefore,  using Eqs.~(\ref{master}) and (\ref{p0n}), one obtains
\begin{eqnarray}\label{tau-phi}
\tau = \left[T^-_1\pi_{1} +T^+_{N-1}\pi_{N-1}\right]^{-1}, \;\; \text{and} \;\;\phi = T^-_1\pi_{1}\tau.
\end{eqnarray}
Similarly, the respective \textit{conditional} MFTs of species $\textsf{A}$ and $\textsf{B}$ (conditioned on the fixation of type $\textsf{A}$ and $\textsf{B}$, respectively) are $\tau^{A}=\left[T^-_{N-1}\pi_{N-1}\right]^{-1}$ and $\tau^{B}=\left[T^+_{1}\pi_{1}\right]^{-1}$. According to Eq.~(\ref{tau-phi}), these quantities are determined once we have obtained $\pi_1$ and $\pi_{N-1}$ from the full expression of the QSD that we now compute.

\begin{figure}
\includegraphics[width=3.4in, height=1.6in,clip=]{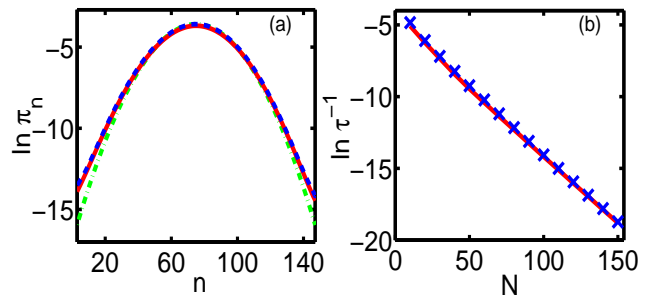}
\caption{{\it (Color online)}.
 (a) $\ln{\pi_n}$
 vs. $n$ (with $N=150$): theoretical predictions [Eqs.~(\ref{fastmode})-(\ref{MPS})] (solid) compared with numerical results (dashed) and with the Gaussian approximation of the QSD (dashed-dotted). (b) $\ln{\tau^{-1}}$  as a function
of  $N$: theoretical
 predictions [Eqs.~(\ref{tau-phi}), (\ref{pi1})-(\ref{MPS})] (solid) and numerical results (symbols). Parameters are $a=0.1$, $b=0.7$, $c=0.6$, $d=0.2$, $w=0.5$ and the system follows the fMP.}
  \label{EPL_fig1}
\end{figure}

The QSD satisfies the quasi-stationary master equation, obtained by substituting Eq.~(\ref{pin}) into (\ref{master}) and neglecting the
exponentially small term $\pi_n/\tau$ (to be verified a posteriori):
\begin{equation}\label{QSME}
T^+_{n-1} \pi_{n-1}+T^-_{n+1} \pi_{n+1} -[T^+_{n}+T^-_{n}]\pi_n=0.
\end{equation}
For $N\gg 1$, we define the  transition rates ${\cal T}_{\pm}(x)= T^{\pm}_n$~\cite{rates} as continuous  functions of $x$,
and treat Eq.~(\ref{QSME}) by employing the WKB ansatz~\cite{kubodykman,EK,AM}
\begin{equation}
\label{ansatz}
\pi_n\equiv \pi_{xN}=\pi(x)={\cal A}\exp[-NS(x)-S_1(x)]\,,
\end{equation}
where  $S(x)$ and  $S_1(x)$ are respectively  the system's action and its amplitude, and ${\cal A}$ is a constant prefactor introduced for convenience. The WKB approximation is here an asymptotic series expansion in powers of $1/N$ based on the exponential ansatz~(\ref{ansatz})~(see, e.g., \cite{kubodykman,EK,AM})~\cite{semicl}. Substituting (\ref{ansatz}) into Eq.~(\ref{QSME}) yields closed equations for $S(x)$ and  $S_1(x)$. To leading order, similarly as in Hamiltonian systems, the action obeys the Hamilton-Jacobi equation $H(x,S')=0$. In this case, the  underlying Hamiltonian is
\begin{eqnarray}
 \label{Hamiltonian}
H(x,p)={\cal T}_+(x)(e^p-1)+{\cal T}_{-}(x)(e^{-p}-1),
\end{eqnarray}
where we have introduced the auxiliary momentum $p(x)=dS/dx$ \cite{kubodykman,EK,AM}. Therefore, to leading order, the ``optimal-path'' followed by the stochastic system, from the metastable state to fixation,  is $p_a(x)=-\ln{[{\cal T}_+(x)/{\cal T}_-(x)]}$, corresponding to the zero-energy trajectory $H(x,p_a)=0$ with non-zero momentum.
The action along $p_a(x)$ is
\begin{equation}\label{Ssingle}
S(x)=-\int^x\ln \left[{\cal T}_{+}(\xi)/{\cal T}_-(\xi)\right]\,d \xi.
\end{equation}
Performing the subleading-order calculations, one obtains
$S_1(x)=(1/2)\ln [{\cal T}_+(x) {\cal T}_-(x)]$~\cite{EK,AM}.
Imposing the normalization of the Gaussian expansion of the QSD (\ref{ansatz}) about $x=x^*$, one finds the constant ${\cal A}$, yielding
\begin{eqnarray}\label{fastmode}
\pi(x) ={\cal T}_+(x^*)\,\sqrt{\frac{S''(x^*)}{2\pi N\,{\cal T}_+(x){\cal T}_-(x)}}\,e^{-N[S(x)-S(x^*)]}\,.
\end{eqnarray}
This expression is valid sufficiently far from the boundaries, where ${\cal T}_{\pm}(x)={\cal O}(1)$ \cite{AM}, and generally leads to a
{\it non-Gaussian} QSD with systematic deviations from the Gaussian approximation near the tails, as illustrated in Fig.~\ref{EPL_fig1}(a).

To obtain the full QSD we  need to match the WKB result (\ref{fastmode}) with the solution of Eq.~(\ref{QSME}) in the vicinity of the absorbing boundaries, where the transition rates can be linearized~\cite{AM}. For instance, near $x=0$,  ${\cal T}_{\pm}(x)\simeq x {\cal T}'_{\pm}(0)$, so Eq.~(\ref{QSME}) yields $(n-1){\cal T}_+^{\prime}(0)\pi_{n-1}+(n+1){\cal T}_-^{\prime}(0)\pi_{n+1}
-n[{\cal T}_+^{\prime}(0)+{\cal T}_-^{\prime}(0)]\pi_n=0$. Its recursive solution is
$\pi_n=(\pi_1/n)(R_0^n-1)/(R_0-1)$, where $R_0={\cal T}_+^{\prime}(0)/{\cal T}_-^{\prime}(0)$.
Matching this expression with the leading order of Eq.~(\ref{fastmode}) in the vicinity of $x=0$ yields
\begin{eqnarray}\label{pi1}
\pi_1\!=\!\sqrt{\frac{N S''(x^*)}{2\pi}}\frac{{\cal T}_+(x^*)\left(R_0-1\right)}{\sqrt{{\cal T}_+^{\prime}(0){\cal T}_-^{\prime}(0)}}e^{N[S(x^*)-S(0)]}.
\end{eqnarray}
A similar analysis at $x\simeq 1$ with $R_1={\cal T}_-^{\prime}(1)/{\cal T}_+^{\prime}(1)$ gives
\begin{eqnarray}\label{pinm1}
\pi_{N-1}\!=\!\sqrt{\frac{N S''(x^*)}{2\pi}}\frac{{\cal T}_+(x^*)(R_1-1)}{\sqrt{{\cal T}_+^{\prime}(1){\cal T}_-^{\prime}(1)}}e^{N[S(x^*)\!-\!S(1)]}.
\end{eqnarray}
Hence, the expressions (\ref{fastmode})-(\ref{pinm1}) provide us with the complete QSD.
\begin{figure}
\includegraphics[width=3.4in, height=2.8in,clip=]{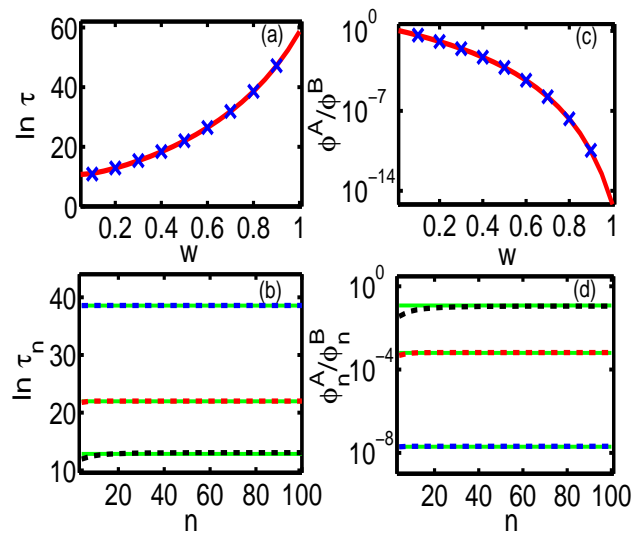}
\caption{{\it (Color online)}. (a) $\ln{\tau}$ vs $w$: theoretical [Eqs.~(\ref{tau-phi}), (\ref{pi1})-(\ref{MPS})] (solid) and numerical results (symbols). (b)
Dependence of $\ln{\tau}$ on the initial number $n$ of $\textsf{A}$'s, for $w=0.2$, $0.5$ and $0.8$ (bottom to top): comparison between theoretical (solid) and numerical (dashed) results. (c) Theoretical [Eq.~(\ref{probfixratio})] (solid) and numerical (symbols) results for the
ratio $\phi^{A}/\phi^{B}$  vs $w$.
(d) Same as in panel (b) for $\phi^{A}/\phi^{B}$ ($w$ grows from top to bottom). Parameters are $a=0.1$, $b=0.7$, $c=0.7$, $d=0.2$, $N=200$ and the system follows the fMP. In the numerical results of (a) and (c), $n$ is chosen sufficiently large so that fixation does not occur  immediately (see text).}
 \label{EPL_fig2}
\end{figure}

\section{Fixation in anti-coordination games}
We now apply the general results obtained in the previous section to study fixation in ACG, when the system follows the fMP.
In this case the action given by Eq.~(\ref{Ssingle}), becomes
\begin{eqnarray}\label{MPS}
S(x)&=&[B/(B-A)-x]\ln [Ax+B(1-x)]\nonumber\\
&+&[D/(C-D)+x]\ln [Cx+D(1-x)],
\end{eqnarray}
where $A=1-w+wa$, $B=1-w+wb$, $C=1-w+wc$, and $D=1-w+wd$~\cite{Claussen}. Provided that $N[S(1)-S(x^*)]\gg 1$, and $N[S(0)-S(x^*)]\gg 1$ (which imposes a lower bound on $w$), the MFTs and fixation probability are obtained from Eqs.~(\ref{tau-phi}) and (\ref{pi1})-(\ref{MPS}) with $T^-_1=T^+_{N-1}\simeq N^{-1}$. These results generalize those obtained  previously  in the limiting cases $Nw\ll 1$~\cite{Traulsen,Hauert-preprint} and $w=1$ (for which $A=a$, $B=b$, $C=c$, and $D=d$)~\cite{Antal}. As  illustrated in Fig.~\ref{EPL_fig1}(b), one finds that  the unconditional MFT asymptotically exhibits an exponential dependence on the population size $N$, $\tau \propto N^{1/2}e^{N(\Sigma-S(x^*))}$, where  the governing exponent  ${\Sigma}\equiv {\rm min}\left[S(0), S(1)\right]$  is readily obtained from (\ref{MPS}). With (\ref{fastmode})-(\ref{pinm1}), this confirms that $\pi_n/\tau$ is indeed exponentially small.
For $0<w<1$, one finds that ${\Sigma}$ increases monotonically with $w$, as shown in Fig.~\ref{EPL_fig2}(a). Here (as in our other figures), the theoretical predictions are compared with the numerical solution of the master equation~(\ref{master}) yielding an excellent agreement.
It also follows from (\ref{tau-phi}),(\ref{pi1})-(\ref{MPS}) that
for $N\gg 1$ and small (but not too small) selection intensity, $N^{-1}\ll w\ll 1$, the conditional MFTs grow exponentially as
$\,\tau^A\sim N^{1/2}e^{Nw(a-c)^2/[2(c-a+b-d)]}$, and $\,\tau^B\sim N^{1/2}e^{Nw(b-d)^2/[2(c-a+b-d)]}$, with $\tau=\tau^A\tau^B/(\tau^A + \tau^B)\sim {\rm min}(\tau^A, \tau^B)$.

As our approach assumes that fixation occurs \textit{after} the metastable state is reached, the
expressions obtained for the MFTs are \textit{independent} of the initial number $n$ of $\textsf{A}$'s, when $n\gg 1$.
This is confirmed in Fig.~\ref{EPL_fig2}(b) where theory and numerical results agree excellently.

The ratio $\phi^A/\phi^B=\phi^{-1}-1$ between the fixation probabilities of the $\textsf{A}$'s  and $\textsf{B}$'s allows to understand the influence of selection and the interplay between selection and demographic stochasticity. Indeed, with Eqs.~(\ref{tau-phi}), and (\ref{pi1})-(\ref{MPS}), our theory  yields
\begin{eqnarray}\label{probfixratio}
\frac{\phi^A}{\phi^B}\!=\!\frac{\pi_{N-1}}{\pi_1}\!=\!\sqrt{\frac{BD}{AC}}\left(\frac{C\!-\!A}{B\!-\!D}\right)
\frac{B^{N\left(\frac{B}{B-A}\right)}D^{N\left(\frac{D}{C-D}\right)}}{A^{N\left(\frac{A}{B-A}\right)}C^{N\left(\frac{C}{C-D}\right)}}\!.
\end{eqnarray}
In Fig.~\ref{EPL_fig2}(c), we show the ratio $\phi^A/\phi^B$  and find a nontrivial exponential dependence on $w$
in excellent agreement with numerical calculations. Contrary to the neutral case $w=0$ (not covered by our theory), where
the ratio of fixation probabilities
strongly depends on the initial number of $\textsf{A}$'s~\cite{Nowak},  Eq.~(\ref{probfixratio})
predicts that $\phi^A/\phi^B$ is independent of the initial condition when the selection strength is finite.
Indeed, the numerical results presented in Fig.~\ref{EPL_fig2}(d) confirm that for $n\gg 1$, the ratio $\phi^A/\phi^B$ coincides with (\ref{probfixratio}) and becomes independent of $n$ when $w$ is nonzero (for $w\ll 1$ the convergence requires $n\sim Nx^*$).

\section{WKB theory and fixation in coordination games} As a further illustration of our
theory, we accurately  compute the fixation probability in CG (\textit{e.g.} stag-hunt
game~\cite{Hofbauer}). Here, the fixed point $x^*$ is a repellor while
$x=0,1$ are attracting, hence there is no metastability and fixation occurs quickly~\cite{Antal}. As a result, with an initial minority of $\textsf{A}$'s, $n<Nx^*$,
the fixation of $\textsf{B}$'s is almost certain, and we are interested in calculating the exponentially small probability $\phi^A_n\equiv \phi^A(x)$ that $\textsf{A}$'s fixate. Such a probability satisfies the following equation~\cite{Antal,Nowak,Hauert-preprint}:
\begin{eqnarray}
\label{CGeq}
T^{+}_n\phi_{n+1}^{A}+T^{-}_n\phi_{n-1}^{A}-[T^{+}_n +T^{-}_n]\phi_n^{A}=0,
\end{eqnarray}
which is the stationary backward master equation of this problem~\cite{Gardiner}, with boundary conditions $\phi_0^{A}=0,\phi_N^{A}=1$.
At this point, it is convenient to introduce the auxiliary quantity
\begin{eqnarray}\label{pn}
 {\cal P}_n \equiv \phi^A_{n+1}-\phi^A_n  \equiv {\cal P}(x),
\end{eqnarray}
which is a normalized PDF peaked at $x^*$. From Eq.~(\ref{pn}), the fixation probability $\phi^A_n$ can be easily obtained, yielding $\phi^A_n=\sum_{m=0}^{n-1}{\cal P}_m$. Substituting Eq.~(\ref{pn}) into Eq.~(\ref{CGeq}), one arrives at a difference equation for the PDF ${\cal P}(x)$ which reads
\begin{eqnarray}
\label{P-eq}
 {\cal T}_{+}(x){\cal P}(x)-{\cal T}_{-}(x){\cal P}(x-1/N)=0.
\end{eqnarray}
 This equation can be treated with the WKB ansatz ${\cal P}(x)={\cal A}_{\rm CG} e^{-N{\cal S}(x)-{\cal S}_1(x)}$.
To leading order one has
${\cal T}_{+}(x)-{\cal T}_{-}(x)e^{{\cal S}'(x)}=0,$ whose solution is
${\cal S}(x)=-S(x)$ [where $S(x)$ is given by Eq.~(\ref{Ssingle})]. In the subleading order, after some algebra, one finds ${\cal S}_1(x)=(1/2)\ln [{\cal T}_+(x)/{\cal T}_-(x)]$.
Normalizing $\sum_{n=0}^{N-1}{\cal P}_n \simeq N\int_{0}^{1} {\cal
P}(x)dx=1$, and assuming a main Gaussian contribution arising
from  $x \simeq x^*$, we find
${\cal A}_{\rm CG}\simeq \sqrt{|S''(x^*)|/(2\pi N)}e^{-NS(x^*)}$.
In the realm of the WKB approach, we have thus obtained an expression of ${\cal P}(x)$ that holds for $0\leq x \leq 1$. [It can be checked that
such a WKB result satisfies Eq.~(\ref{P-eq}) also near the absorbing boundaries~\cite{long-paper}.] From the expression of
${\cal P}(x)$, the fixation probability thus reads
\begin{eqnarray}\label{cdfcoord}
\phi^A_n=\sqrt{\frac{|S''(x^*)|}{2\pi N}}\sum_{m=0}^{n-1} \sqrt{T^-_m/T^+_m}\;e^{N[S(m/N)- S(x^*)]}.
\end{eqnarray}

\begin{figure}
\begin{center}\includegraphics[width=3.4in, height=1.7in,clip=]{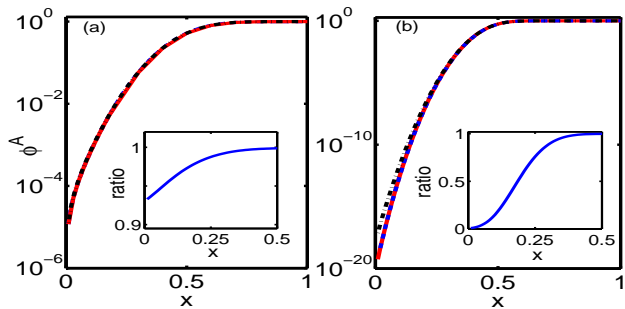}
\caption{{\it (Color online)}. The fixation probability $\phi^{A}(x)$
for the fMP process: theoretical result (\ref{cdfcoord}) (solid),
numerical calculations (dashed) and
  FPA (dash-dotted), with $a=4, b=0.2, c=0.3, d=3.8, N=100$. Insets:
ratio between theoretical results and those of the FPA, see text.
(a) For $w=0.1$, $Nw^2=1$ and all curves agree well, with an error of
about $7\%$ in the predictions of the FPA for $x\to 0$.
(b) For $w=0.75$, $Nw^2 \gg 1$, the curve obtained from the FPA
systematically deviates from the others and yields exponentially large
errors.
} \label{EPL_fig3}
\end{center}
\end{figure}

\begin{figure}
\includegraphics[width=3.4in, height=1.6in,clip=]{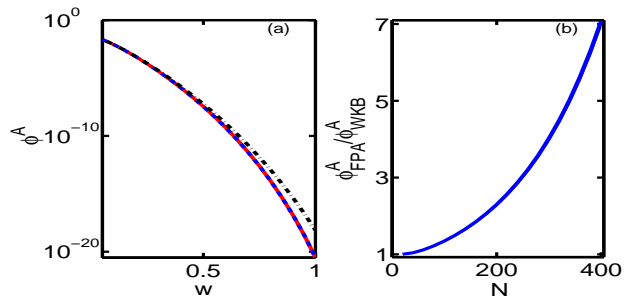}
\caption{{\it (Color online)}. (a) Fixation probability $\phi^{A}(x)$
as function of $w$: theoretical result (\ref{cdfcoord}) (solid),
numerical calculations (dashed) and FPA (dash-dotted), for $a=1$,
$b=0.2$, $c=0.3$, $d=0.8$, and $N=200$.
(b) Ratio between the predictions of the FPA and those of our theory
vs $N$, for $w=0.25$, $a=4$, $b=0.2$, $c=0.3$, and $d=3.8$. The
results of the FPA deteriorate when both $w$ and $N$ increase.
In (a) and (b), $n=10$ thus $x=10/N$, and
the system follows the fMP.} \label{EPL_fig4}
\end{figure}

Of special interest is the limit of $n\ll Nx^*$ corresponding to the fixation of a few mutants in a sea of wild-type individuals~\cite{Nowak}. In this case, it can be shown that Eq.~(\ref{cdfcoord}) can be well approximated by
$\phi^A(x) \simeq  {\cal P}(x)/[e^{S'(x)}-1]$ when $w={\cal O}(1)$, while for small selection $N^{-1}\ll w \ll 1$, $\phi^A(x) \simeq \sqrt{N|S''(x^*)|/(2\pi)}\int_{0}^{x} dy \;e^{N[S(y)-S(x^*)]}$~\cite{long-paper}.
A comparison between theory (\ref{cdfcoord}) and numerical results, using $S(x)$ from Eq.~(\ref{MPS}), is shown in Figs.~\ref{EPL_fig3} and \ref{EPL_fig4}(a) and an excellent agreement is observed.

\vspace{0.2cm}

The WKB theory presented in this section (as well as that dedicated to the ACG) is valid as long as $w\gg N^{-1}$. For small selection intensity, $w\ll 1$~\cite{sel}, the fixation probability is often computed using the FPA (or diffusion approximation)~\cite{Traulsen,Kimura,Hauert-preprint}, usually considered within the linear noise approximation~\cite{Gardiner}. Thus, for $N^{-1}\ll w\ll 1$ (and $N\gg 1$), the predictions of the WKB and FPA approximations can be compared (together with results of numerical  simulations) to determine their respective domains of validity.

For the purpose of  comparison, it is convenient to rewrite both WKB and FPA predictions in the following form: $\phi^A(x) \simeq \Psi(x)/\Psi(1)$, where
$\Psi(x)=\int_0^x dy\, e^{-\int_0^y dz \, \Theta(z)}$. From Eq.~(\ref{cdfcoord}), one finds that for the WKB approach the exponent reads
$\Theta_{\rm WKB}(z)=N\ln{\left[{\cal T}_+(z)/{\cal T}_-(z)\right]}$, whereas for the FPA one has the exponent
$\Theta_{\rm FPA}(z)=2Nz\left[{\cal T}'_+(x^*) - {\cal T}'_-(x^*)\right]/\left[{\cal T}_+(x^*) + {\cal T}_-(x^*)\right]$~\cite{Gardiner}.
Hence,  it can be shown that in the vicinity of $x=x^*$, $\Theta_{\rm WKB}(x) - \Theta_{\rm FPA}(x)\sim Nw^2(x-x^*)^2$~\cite{long-paper}.
Therefore, while  the WKB result (\ref{cdfcoord}) is accurate for any finite value of $w$ [as shown in Fig.~\ref{EPL_fig4}(a)],
the FPA is unable to account for fixation and yields exponentially large errors when $w\gtrsim  N^{-1/2}$.
In fact, the predictions of the FPA (within linear noise approximation) are accurate only
when the selection intensity satisfies $w \ll N^{-1/2}$, which is a more stringent condition than $w\ll 1$. This is illustrated in Figs.~\ref{EPL_fig3} and \ref{EPL_fig4} which display a comparison between our predictions and those of the FPA for various values of $w$ and $N$.

\section{Conclusion}
We have studied fixation in evolutionary games under non-vanishing selection and elucidated the nontrivial relation between selection intensity and effects of demographic fluctuations. This has been achieved by generalizing a recent WKB-based theory to account for multiple absorbing states. This approach naturally accounts for non-Gaussian behavior and allows an accurate treatment of large fluctuations.
In the framework of models of cooperation dilemmas, we have analytically computed the QSD (shape of the metastable PDF),
MFTs and the fixation probabilities \textit{beyond} the weak selection limit.
While it does not cover the $w\to 0$ limit (where the FPA holds), our theory agrees excellently with numerical simulations
over a broad range of finite selection strength ($0<w\leq 1$), where  the FPA generally fails.
For concreteness, our approach has been illustrated for two classes of (formally solvable) $2 \times 2$ games, but is
neither restricted to linear payoffs nor to a specific choice of the transition rates~\cite{long-paper}. Importantly,  our theory  can be adapted to study evolutionary processes for which there is no rigorous analytical treatment (e.g. $3 \times 3$ games~\cite{Hofbauer}) and help generalize the
concept of evolutionary stability.

\vspace{0.3cm}

{\bf Acknowledgments:} We thank Baruch Meerson for a useful discussion.


\begin{thebibliography}{99}
\bibitem{rates} For simplicity we neglect self-interaction ${\cal O}(1/N)$ terms and keep only the leading-order ${\cal O}(1)$ terms in
$\Pi_{A,B}(n)$, and as a result, also in $T^{\pm}_n$ which appear in Eq.~(\ref{master}).
\bibitem{choice} Other choices for the (positive definite and ``well-behaved'') functions $\chi^{\pm}(n)$ are possible, as long as these obey the necessary condition $\chi^+(Nx^*)=\chi^-(Nx^*)$ to ensure the recovery of the mean-field properties~(\ref{RD}) in the limit $N\to \infty$~\cite{Claussen06,long-paper}.
\bibitem{sel}  It has to be noted that the condition $w\ll 1$ describes a much wider regime than that of the weak-selection limit defined by $Nw \ll 1$.
\end{thebibliography}
\end{document}